\documentclass[twocolumn,twoside,slac_two]{revtex4}
\usepackage{graphicx}
\usepackage{fancyhdr}
\begin{document}

\title{Constraining Pulsar Emission Physics through Radio/Gamma-Ray Correlation of Crab Giant Pulses}

%

\author{A.V. Bilous}
\affiliation{Department of Astronomy, University of Virginia, PO Box 400325, Charlottesville, VA 22904}

\author{V.I. Kondratiev}
\affiliation{Netherlands Institute for Radio Astronomy (ASTRON), Postbus 2, 7990 AA Dwingeloo, The Netherlands}
\author{M.A. McLaughlin, M. Mickaliger, D.R. Lorimer}
\affiliation{Department of Physics, West Virginia University, Morgantown, WV 26506}
\author{S.M. Ransom}
\affiliation{NRAO, Charlottesville, VA 22903}
\author{M. Lyutikov}
\affiliation{Department of Physics, Purdue University, 525 Northwestern Avenue, West Lafayette, IN 47907-2036}
\author{B. Stappers}
\affiliation{Jodrell Bank Centre for Astrophysics, School of Physics and Astronomy, The University of Manchester, Manchester M13 9PL, UK}
\author{G.I. Langston}
\affiliation{NRAO, Green Bank, WV 24944}

\begin{abstract}
To constrain the giant pulse (GP) emission mechanism and test the model of Lyutikov~\cite{lyutikov2007} 
of GP emission, we are carrying out a campaign of simultaneous observations of the Crab pulsar between 
$\gamma$-rays (Fermi) and radio wavelengths. The correlation between times of arrival of radio GPs and
high-energy photons, whether it exists or not, will allow us to choose between different origins of GP 
emission and further constrain the emission physics. Our foremost goal was testing whether radio GPs are 
due to changes in the coherence of the radio emission mechanism, variations in the pair creation
rate in the pulsar magnetosphere, or changes in the beaming direction. Accomplishing this goal requires 
an enormous number of simultaneous radio GPs and $\gamma$-photons. Thus, we organized a radio observations 
campaign using the 42-ft telescope at the Jodrell Bank Observatory (UK), the 140-ft telescope, and the 100-m 
Robert C. Byrd Green Bank Telescope (GBT) at the Green Bank Observatory (WV). While the observations with 
the two first ones are ongoing, here we present the preliminary results of 20~hrs of observations with the 
GBT at the high frequency of 8.9~GHz. These particular observations were aimed to probe the model of GP emission 
by Lyutikov~\cite{lyutikov2007} which predicts that GPs at frequencies $> 4$~GHz should be accompanied
by $\gamma$-ray photons of energies of 1-100~GeV.
\end{abstract}

\maketitle

\thispagestyle{fancy}


\section{Introduction}

The Crab pulsar was discovered by Staelin \& Reifenstein in 1968~\cite{staelin1968} by its remarkably
bright giant pulses (GPs). These relatively rare bursts of intense radio emission last a few nanoseconds 
to a few microseconds, and are clearly a special form of pulsar radio emission~\cite[e.g.~][]{knight06,popsta07}.
GPs generally occur only in certain narrow ranges of pulse phase that are often coincident with
pulses seen at X-ray and $\gamma$-ray energies~\cite{Lundgren}. Popov et al.~\cite{pop06}
propose that all radio emission from the Crab (except for that in the precursor) is composed
entirely of GPs, consistent with the alignment of the GP and high-energy components seen in other
GP pulsars~\cite{cusumano2003, knight06}.

Crab pulsar shows pulsed emission across the entire electromagnetic spectrum (see Fig.~\ref{fig:prof}, left),
reflecting different radiation processes in pulsar magnetosphere~--- from coherent curvature or
synchrotron (radio) to incoherent synchrotron (optical and X-ray) and incoherent curvature ($\gamma$-ray).
Similar to other sporadic, variability phenomena in pulsar radio emission, represented by nulling
pulsars~\cite[e.g.~][]{herfindal2009}, intermittent pulsars~\cite{kramer06}, and rotation radio transients~\cite{mam06}, 
GP emission can be due to changes in the
coherence of the radio emission, variations in the pair creation rate in the magnetosphere, or changes
in the beaming direction.
If GP phenomenon is due to changes in the coherence of the radio emission mechanism, then one would expect little
correlation of the radio GPs with the high-energy emission.  However, if the GPs are due to
changes in the actual rate of pair creation in the pulsar magnetosphere, one would expect an increased flux 
at high energies at the
time of the GPs. Similarly, because the radio GP and $\gamma$-ray components are aligned, one expects that 
they come from the same place in the pulsar magnetosphere. Therefore, if a GP occurs from a beam direction 
alteration, one would expect to see an increase also in the high-energy flux.

The attempt to carry out simultaneous radio/$\gamma$-ray observations (50-220~keV energy range of CGRO/OSSE)
and correlate time of arrivals (TOAs) of GPs was undertaken before by Lundgren et al.~\cite{Lundgren}. They did not
have enough sensitivity to correlate TOAs for single events and averaged their data in 2-min intervals.
They were only able to put an upper limit on $\gamma$-ray flux increase of a factor of 2.5 concurrent
with radio GPs. This suggested that GP mechanism is largely based on changes in coherence and not changes
in pair production rates or beaming.
Yet, Shearer et al.~\cite{shearer2003} performed simultaneous radio/optical observations
of the Crab pulsar, and they found a weak correlation, that optical pulses coincident with radio GPs were
on average 3\% brighter than others. 
In contrast to the Lundgren et al. work, this observation clearly points to
variations in magnetospheric particle density as the cause of the radio giant pulses.

Also, Lyutikov~\cite{lyutikov2007} proposed a more specific, quantitative model of GP emission
in which Crab GPs are generated on closed magnetic  field lines near the light cylinder via anomalous cyclotron
resonance on  the ordinary mode. One clear prediction of this model is that radio GPs (at least those at
radio frequencies $> 4$~GHz) should be accompanied by $\gamma$-ray photons,
as the high energy beam is expected to produce curvature radiation at energies $ \sim  \hbar \gamma^3 \Omega \sim  1$--100~GeV,
depending on the exact value of the Lorentz factor $\gamma$. These energies fall perfectly into the 
energy range of the Fermi mission, and so this hypothesis can also be tested through
high-frequency radio observations concurrent with Fermi.

In Sections~\ref{radioobs} and \ref{fermidata} below we describe performed radio observations and Fermi data
used in further analysis. Section~\ref{results} presents the obtained preliminary results. We describe the correlation
analysis between radio GPs and Fermi photons in Section~\ref{correlation}. Conclusions are made in
Section~\ref{conclusions}.

\section{Radio observations}\label{radioobs}

The radio observations were carried out during the September 2009 with the 100-m Robert
C. Byrd Green Bank Telescope (GBT) using the new Green Bank Ultimate Pulsar Processor (GUPPI)
at a frequency of 8.9~GHz. The total bandwidth of 800~MHz
was split into 256 frequency channels, and the total intensity was recorded with the sampling
interval 2.56--3.84 $\mu$s. There were 10 observing sessions for a total of $\sim$20~hrs.

\begin{figure}
\includegraphics[scale=0.43]{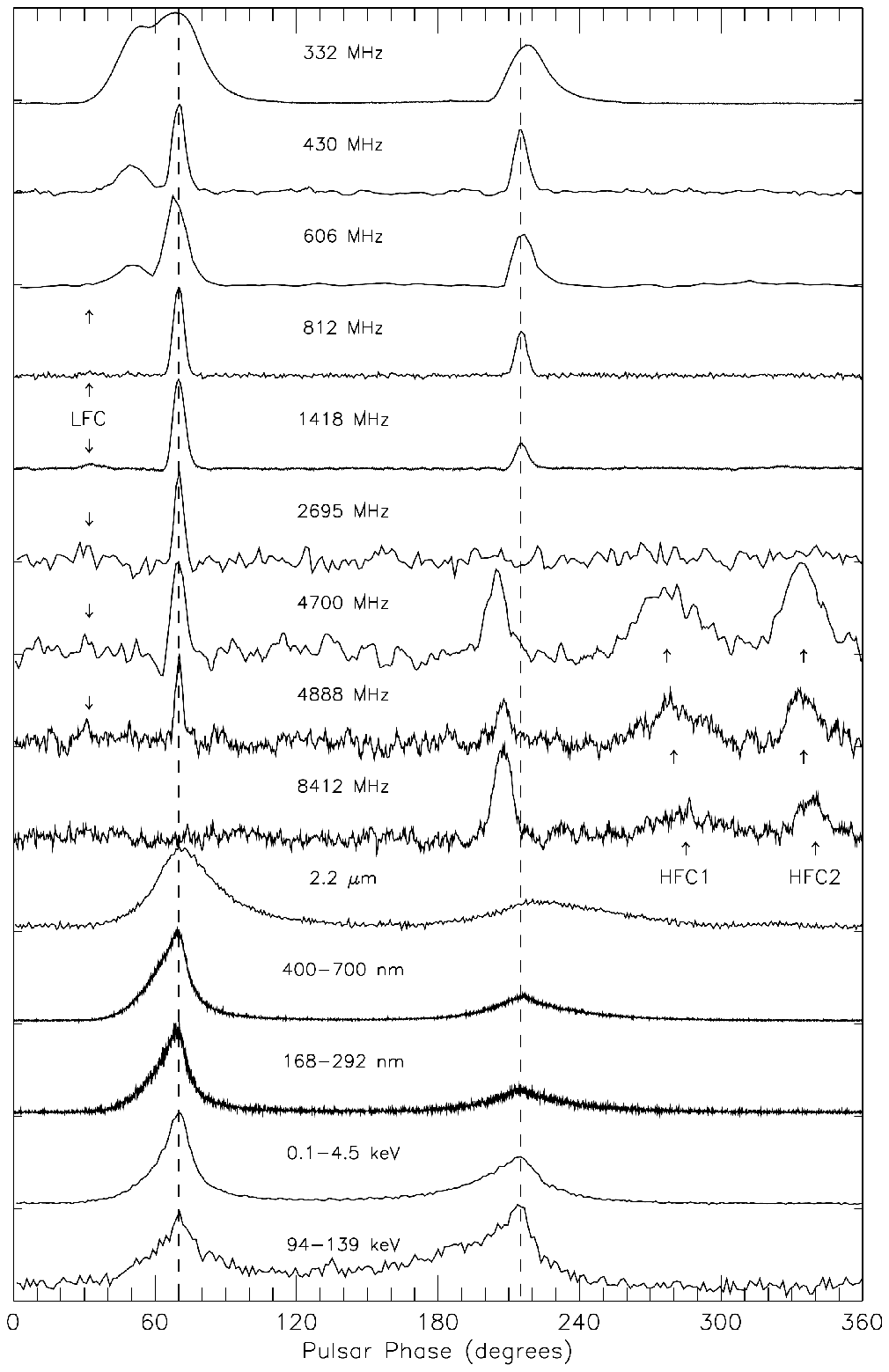}\includegraphics[scale=0.5]{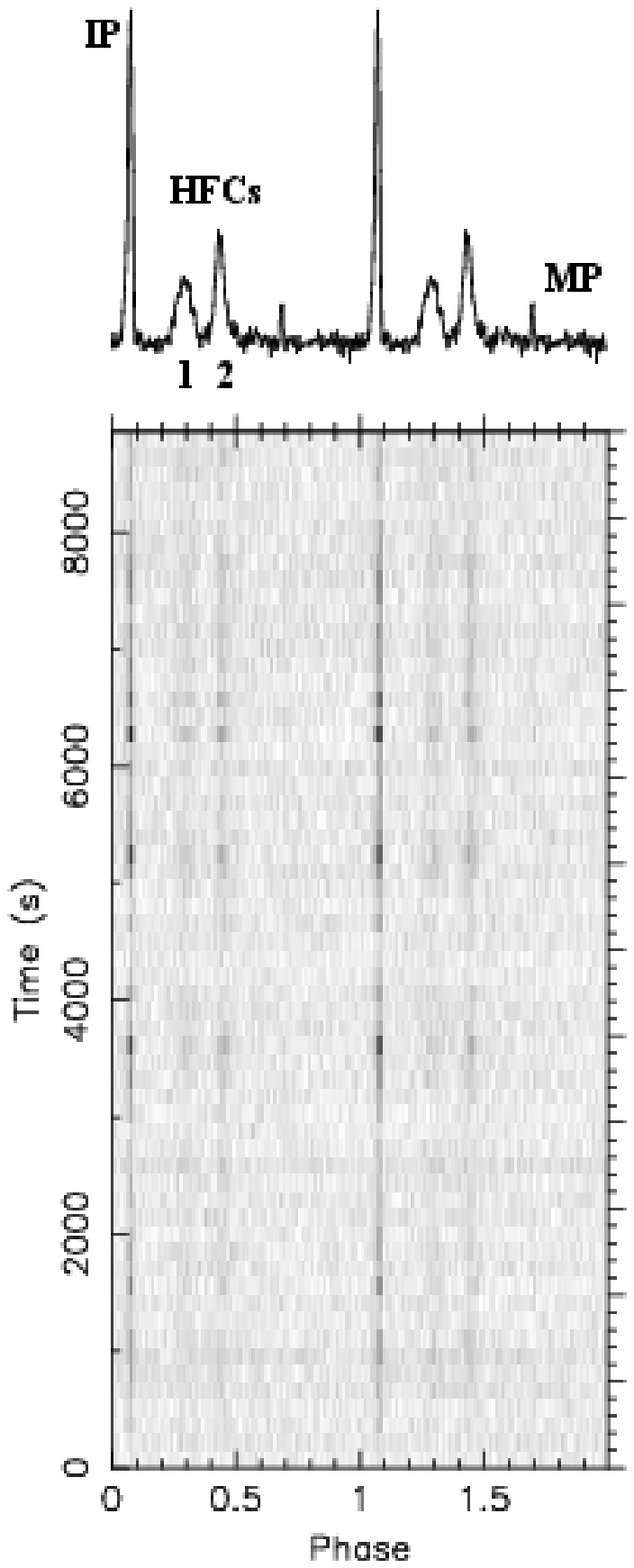}%
\caption{{\it Left.} The average profile of the Crab pulsar from radio to $\gamma$-rays (from the paper of 
Moffett \& Hankins~\cite{moffett1996}). {\it Right.} Average Crab pulsar radio profile for the GBT session on Sep 25, 2009.\label{fig:prof}}
\end{figure}

The raw data from every session were dedispersed with the current DM of the Crab Pulsar using
the PRESTO package, and searched for all the single-pulse events. The lists of events were
presented in TEMPO format and converted to barycentric reference frame for further analysis with
Fermi data. Figure~\ref{fig:prof}, right shows the average profile (top) of the Crab Pulsar together with the
subintegrations during the course of observations for the most fruitful session on Sep~25. The
interpulse (IP) and high-frequency components (HFCs) are clearly seen, with the weak peak after HFC2
being the main pulse (MP). 

\section{Fermi data}\label{fermidata}

For each observing radio session we extracted LAT data for diffuse class of events.
Photons with energies above 100~MeV and with the angular separation 
$\theta < \mathrm{Max}(1.6 - 3\lg(E/1000~\mathrm{MeV}),1.3) ^\circ$ from the nominal pulsar position were 
selected (e.g.~\cite{grondin2009}). Total number of photons per 8~hrs of simultaneous observations is 70 (11 above 1~GeV).
Photons were barycentered with TEMPO2 using the same ephemeris as for radio GPs. 

\section{Results}\label{results}

The number of detected radio GPs stronger than $7\sigma$ and $\gamma$-photons with the energies above 100~MeV
during contemporaneous time for each observing session is given in the Table~\ref{table}.
Taking into account the contribution from the Crab nebula, the 1$\sigma$ sensitivity in our
observations was about 480~mJy. In total, we found more than 85,000~GPs stronger than 7$\sigma$, with
about simultaneous 39,450~GPs with Fermi. The corresponding total number of photons detected is 70.

\begin{table}[t]
\begin{center}
\caption{Observations summary. The columns are as follows: the day in September when observations were
happened; $T_\mathrm{radio, \gamma}$, the duration of radio session simultaneously with Fermi; $N_\mathrm{GPs}$,
the number of detected GPs stronger than $7\sigma$ during contemporaneous time with Fermi; $N_\gamma$,
the number of Fermi photons occurred with energy $> 100$~MeV.}
\begin{tabular}{c c c c c c c c c c}
\hline
Day  & $T_\mathrm{radio, \gamma}$  & $N_\mathrm{GPs}$ & $N_\gamma$ \\
(Sep~2009) & (s)  &  ($>7\sigma$) & ($>100$~MeV)  \\ [0.5ex]
\hline
12 & 1551 & 4 & 7 \\
14 & 3304 & 4166 & 11 \\
19 & 3077 & 4096 & 7 \\
20 & 1779 & 1069 & 2 \\
21 & 1801 & 154 & 2 \\
22 & 3871 & 1567 & 6 \\
23 & 4710 & 9764 & 10 \\
24 & 1260 & 264 & 1 \\
25 & 7446 & 18299 & 19 \\
28 & 2795 & 65 & 5 \\
\hline
Total: & 31594 & 39448 & 70
\end{tabular}
\label{table}
\end{center}
\end{table}

\begin{figure*}[t]
\centering
\includegraphics[scale=0.5]{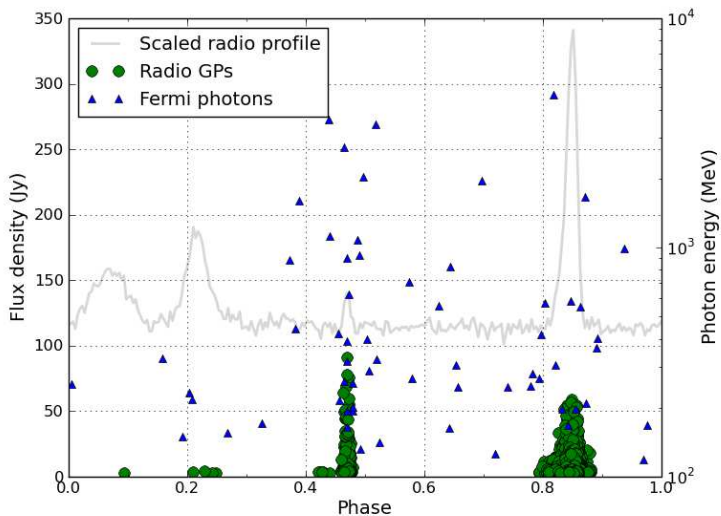}
\caption{Flux density of GPs and energy Fermi photons vs. the phase
of their occurrence within the Crab period. The radio profile from
one of the GBT sessions is shown in the background (gray).}
\label{fig:profenergy}
\end{figure*}

\begin{figure*}[t]
\centering
\includegraphics[scale=0.5]{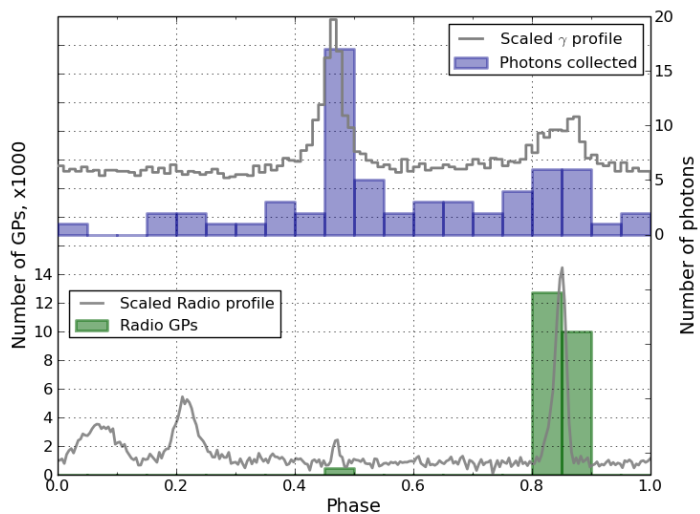}
\caption{Histograms of GPs and Fermi photons during the simultaneous time. For illustrative purposes radio 
and gamma scaled profiles are shown (gray). Scaled radio profile is from GBT session on Sep 25, 2009, 
and $\gamma$-profile is Fermi profile accumulated during Sep, 2009.}
\label{fig:profhist}
\end{figure*}

Figures~\ref{fig:profenergy} and \ref{fig:profhist} present the flux density of radio GPs and energy of Fermi photons
and their histograms
versus the phase of their occurrence within the Crab period. It is evident that most of GPs occurred at the phases 
of the MP and IP. In agreement with Hankins \& Eilek~\cite{hankins2007}  radio GPs occur more frequently 
at the phases of IP, while GPs in MP are stronger. Due to small photons sample, no definitive conclusion 
can be put forward, whether there is an average increase in counts rate in Fermi profile during the events of radio GPs.
Several pulses, either giant pulses or regular single pulses, were also detected
at the phases of HFCs and even MP precursor. Though such pulses were reported by Jessner et al.~\cite{jessner2005} as well,
their GP nature is to be checked, and they could only represent strong single pulses of regular emission.

\begin{figure*}[t]
\centering
\includegraphics[scale=0.5]{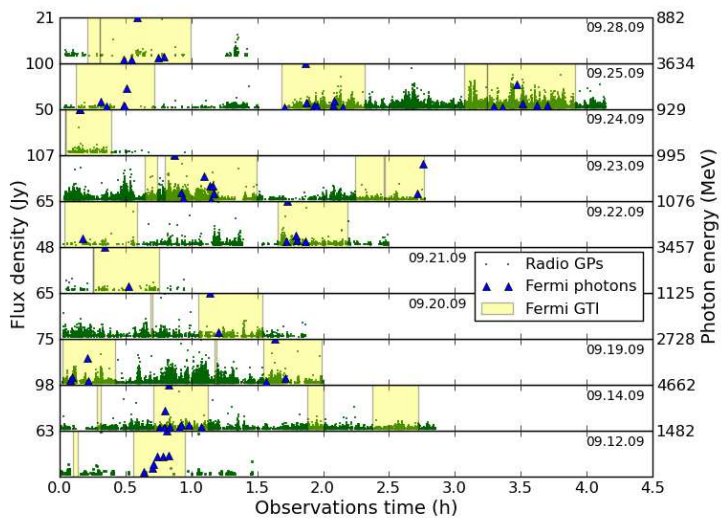}
\caption{Time series of radio GPs and Fermi photons during 10
observing sessions. The values on Y-axes shows either the maximum
flux density in radio (left) or the maximum photon energy (right) in the
corresponding subplot.}
\label{fig:burst}
\end{figure*}

At high radio frequency of 8.9~GHz GPs occur in bursts or clumps due to scintillations 
with a characteristic scintillation time of about 10-20 min. However, scintillations can mask the intrinsic
flux variability inherent to the pulsar. Hence, if so and there is a correlation between time of arrivals
of radio GPs and high-energy photons, one would expect Fermi photons also come within clumps of GPs.
Figure~\ref{fig:burst} shows the time series of radio GPs and Fermi photons for every observing session.
Apparently photons do also have a tendency to occur in clumps, and in some cases during the increase of
the flux in radio. The latter, however, does require an additional analysis.

\subsection{Correlation Analysis}\label{correlation}

For every photon we searched for a GP within the set of different time windows between 1~ms and 10~s. 
Since the rate of GPs varies significantly from session to session (see Fig.~\ref{fig:burst}), 
the search was performed for each day separately. The results are presented in the Table~\ref{table2}.
It is obvious, that the higher the average GP rate and/or length of correlation time window, the
more matches between radio GPs and Fermi photons.

\begin{table}[t]
\begin{center}
\caption{Correlation results between radio GPs and Fermi photons for each
observing session. Column 1 shows the day in September 2009 when observation
was happened, column 2 gives the average GP rate during the session, and 
columns 3--7 give the number of matches between radio GPs and Fermi photons
for different time windows from 1~ms to 10~s.}
\begin{tabular}{ccccccc}
\hline
Day  & GP rate & \multicolumn{5}{c}{Time window} \\
\cline{3-7}
(Sep~2009) & (s$^{-1}$) & 1~ms & 10~ms & 100~ms & 1~s & 10~s \\
\hline
12 & 0.0026 & -- & -- & -- & -- & -- \\
14 & 1.26   & 1  & 1  & 4  & 9  & 10 \\
19 & 1.33   & -- & -- & 2  & 5  & 7  \\
20 & 0.6    & -- & -- & -- & -- & 2  \\
21 & 0.0855 & -- & -- & -- & -- & 1  \\
22 & 0.4    & -- & -- & -- & 1  & 4  \\
23 & 2.073  & -- & 1  & 4  & 7  & 9  \\
24 & 0.21   & -- & -- & 1  & 1  & 1  \\
25 & 2.458  & -- & -- & 6  & 13 & 15 \\
28 & 0.023  & -- & -- & -- & 2  & 2  \\
\hline
\end{tabular}
\label{table2}
\end{center}
\end{table}

To estimate the probability $P$ of getting the same 
number of matches by pure chance (i.e. when there is no true correlation), we performed a Monte-Carlo simulation 
assigning a random time of arrival for each $\gamma$-photon for every observing session. Then, we calculated
the number of matches between real radio GPs and simulated photons and repeated this procedure 
$N_\mathrm{sim} = 10,000$ times.
Knowing the real number of matches $k$ for every observing session and every time window, we calculated
$P$ as $N^k / N_\mathrm{sim}$, where $N^k$ is the number of performed simulations with $k$ matches.

\begin{figure*}[t]
\centering
\includegraphics[scale=0.5]{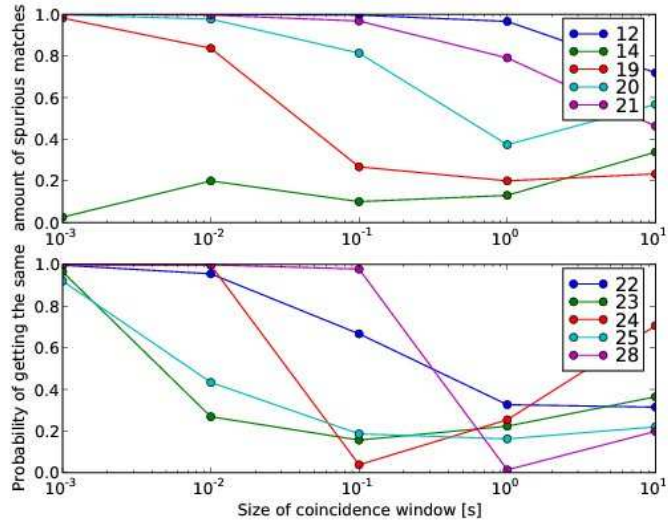}
\caption{The probability that the number of matches between GP and $\gamma$- photon within given time window is due 
to pure chance (no correlation). Each line corresponds to different observation date. The simulation was performed 
for each observing session separately due to different rates of radio GPs.} \label{fig:prob}
\end{figure*}

The results of this simulation are presented on Figure \ref{fig:prob}. It is clear that the probability of 
getting the recorded number of matches between GPs and $\gamma$-photons by pure chance is very high,
and, thus, the presence of intrinsic correlation is still under the question. 


\section{Conclusions}\label{conclusions}

\textit{No obvious correlation was found between Fermi photons of energies $>$ 100~MeV and radio 
giant pulses at the frequency of 8.9~GHz.}

Due to a small number of contemporaneous photons, no definitive inference can be made about an average 
increase in counts rate in Fermi profile during the events of radio GPs. Current preliminary results 
indicate that fraction of photons closely accompanying high-frequency GPs is certainly less than 10\%.
In order to estimate this value more precisely, one needs to continue
simultaneous observations, accumulating more photons and registering more GPs. Such a campaign of Fermi/radio
observations using 140-ft telescope at Green Bank Observatory (WV) and 42-ft telescope at Jodrell Bank Observatory
at low frequencies is ongoing now. Non-correlation, if indeed true, may favor for coherence change as a reason for GP
emission rather than variations in pair creation rate in the pulsar magnetosphere, or changes in beaming direction. In
particular, the model of Lyutikov~\cite{lyutikov2007} predicts correlation between high-energy photons and radio GPs at
frequencies 4-10~GHz. Again, if radio and $\gamma$-rays are indeed not correlated, then model either has to be tweaked,
or such correlation exists only for very high energy photons $\gtrsim 100$~GeV. At this energy range Fermi, though sensitive, 
can not provide extensive sample of photons in reasonable time for correlation, so Cherenkov detectors are much 
more promising. 

Simple identification of time of arrivals of photons shows that they also come in groups similarly to radio GPs. 
Though in case of GPs this is caused by scintillations, they could potentially mask the flux variability intrinsic 
to the Crab pulsar as well. Then, we would expect groups of photons to concentrate with fraction of stronger GPs. 
And in fact, we do see the tendency of some groups of photons to cluster around strong GPs (Fig.~\ref{fig:burst}).
Using the current dataset, we are planning to do a more thorough analysis of TOA correlation between individual
photons and radio GPs. There could be a time delay between time of arrival of Fermi photon and corresponding GP
due to, for instance, different travel paths in the pulsar magnetosphere, or non-simultaneity in emitting radio and
gamma-ray. We will introduce the delay between photon and GP time series and run the correlation analysis for many
such trials. Also, $\gamma$-photons may accompany only the brightest GPs or the clump of giant pulses as a whole.

\bigskip 

\begin{acknowledgments}
The National Radio Astronomy Observatory is a facility of the National Science Foundation operated under
cooperative agreement by Associated Universities, Inc. 
\end{acknowledgments}

\bigskip 


\end{document}